# Tracking the Digital Footprints to Scholarly Articles from Social Media

Xianwen Wang*, Zhichao Fang and Xinhui Guo

WISE Lab, Faculty of Humanities and Social Sciences, Dalian University of Technology, Dalian 116085, China.

* Corresponding author.

Email address: xianwenwang@dlut.edu.cn; xwang.dlut@gmail.com

Website: xianwenwang.com

**Abstract:**

Scholarly articles are discussed and shared on social media, which generates altmetrics. On the opposite side, what is the impact of social media on the dissemination of scholarly articles and how to measure it? What are the visiting patterns? Investigating these issues, the purpose of this study is to seek a solution to fill the research gap, specifically, to explore the dynamic visiting patterns directed by social media, and examine the effects of social buzz on the article visits. Using the unique real referral data of 110 scholarly articles, which are daily updated in a 90-day period, this paper proposes a novel method to make analysis. We find that visits from social media are fast to accumulate but decay rapidly. Twitter and Facebook are the two most important social referrals that directing people to scholarly articles, the two are about the same and account for over 95% of the total social referral directed visits. There is synchronism between tweets and tweets resulted visits. Social media and open access are playing important roles in disseminating scholarly articles and promoting public understanding science, which are confirmed quantitatively for the first time with real data in this study.

# Introduction

With the booming of social media in the last decades, sharing scholarly materials on social web is becoming increasingly popular. Due to the advantages that social media have shown, an efficient channel for scientific communication and dissemination is established, and further strengthened by open access (Wang, et al., 2015). The spread of research results in social media not only enhances the interaction between scientific community and the public, but also provides a wealth of data for gaining new insights into the research activities. Based on the social media data of scholarly articles, altmetrics, which are defined as the alternative or complement to traditional metrics, were designed and proposed (Prime, et al., 2010; Prime and Hemminger, 2010).

Altmetrics mainly record the spread of scholarly materials in digital libraries, reference management systems, social networks and media, including both general and academic (Torres-Salinas, et al., 2013), different from usage metrics, which focus on the usage situations of scholarly articles in various forms, like view, download, visit, save, etc. Making a distinction between these two terms could avoid unnecessary confusion and misunderstandings (Glänzel and Gorraiz, 2015), and make for better understanding of the interaction between the social media spreading and article usage.

According to the mission and original intention of altmetrics, most recent studies turned to make comparisons between altmetrics and conventional metrics. Researchers made a lot of efforts to seek definitive evidence for or against the applicability and validity of altmetrics through calculating the correlations between different kinds of altmetrics and citations (Thelwall, et al., 2013; Haustein, et al., 2013; Bornmann, 2014a, b; Costas, et al., 2014; Sotudeh, et al., 2015). Due to the differences of data sources and disciplinary fields, their results are distinct, and even conflicting. It is hard to claim that altmetrics are substitutes for conventional indicators, but to be sure is that altmetrics could tell a series of stories that conventional indicators could hardly tell, especially the societal impact of scientific results (Bornmann, 2014a, b). Therefore, altmetrics are supposed to be applied to research assessment and funding scheme evaluation from a new perspective (Dinsmore, et al., 2014; Thelwall, et al., 2016). The spread of scholarly articles on social media offers a new approach to make evaluations on scientific results through the form of altmetrics indicators. Even though there are always disputes and criticisms along with their development due to the lack of authority and credibility as performance measures (Cheung, 2013), altmetrics demonstrate their value as data recorders. The altmetrics data with specific information are often used to analyze spreader attributes, spreader behaviors or other related issues (Bornmann and Haunschild, 2015; Mohammadi, et al., 2015).

But as to scholarly articles, what impacts will the social media spreading bring to themselves? Based on the relatively high correlation between some specific altmetrics, e.g., Twitter, with citations, Eysenbach (2011) speculated that the spread of articles in Twitter may increase their cited times. Similarly, Allen et al. (2013) reported that social media release of a research article increases the number of people who view or download that article. However, correlation, after all, does not necessarily imply causation. The impact that social media spreading makes on scholarly articles still remains in the speculation. How to measure and describe the impact? Because of the limitation of the data availability, this research question has never been answered ever.

In our previous study (Wang, et al., 2016), the referral data collected from *PeerJ* are used to track the digital footprints to scholarly articles. We found that social network platforms are top referral sources followed by bookmark or typed URL and general search engines, social media bring lots of visitors and make a great effect on the usage of scholarly articles. Further, which specific social media platform plays the biggest role in referral? What is the temporal trend of the effect produced by social media on the usage of scholarly articles? Is there any meaningful fluctuation in the trend? And what is behind the fluctuation? We will answer these questions in this study.

## Data and methods

*Peer*J is an open access, peer-reviewed, scholarly journal, which considers and publishes research articles in the Biological and Medical Sciences. It published its first articles on February 12, 2013 and received its first (partial) Impact Factor of 2.1 in June 2015.

*PeerJ* includes article level metrics on all of the article pages. The article level metrics enable people to see the number of visitors, views and downloads an article has received, and also links to the referral source of visitors and are updated daily, as Fig. 1 shows. The *PeerJ* article level metrics, especially the referrals data make it unusual and different from the metrics data provided by other publishers, for example, metrics provided by PLOS and Nature only include the article views data and altmetrics data (Wang, et al., 2014). Using this kind of fresh metrics data, we could track the digital footprints to scholarly articles and explore people's visiting patterns from a novel perspective.

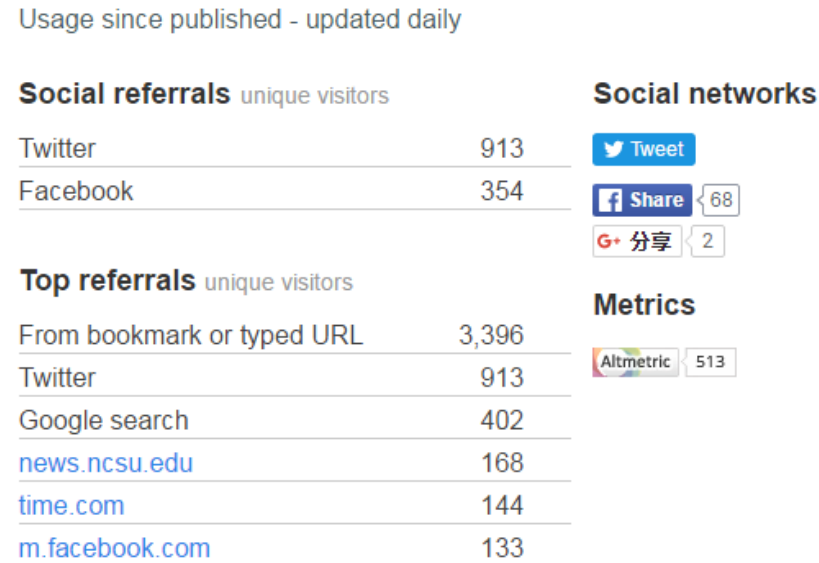

**Fig. 1** Article usage data provided by PeerJ

As shown in Figure 1, the *PeerJ* article level metrics panel consists of three parts, which are the Social referrals at the top left of the panel, Top referrals at the bottom left of the panel, and Social networks and Metrics at the right of the panel. The part of Social referrals shows five mainstream social media platforms where visitors from, which are Facebook, Twitter, LinkedIn, Reddit, and Google+, if any. The part of Top referrals actually displays the visitors from every referral, when the part of Social networks and Metrics reveals the times of being shared to every social media platform and the altmetric score calculated by altmetric.com.

Our data in this study are harvested from the website of *PeerJ* journal, including the metadata and article metrics data, when the article metrics data are from the *PeerJ* article level metrics panel. We have been collecting data since January 22, 2016 and updating the data every day. We check the journal for the new articles published each day and harvest the daily visiting data of each article. In this paper, the daily visiting data from January 22, 2016 to April 20, 2016 are used, which is 90 days in all. To ensure that all the articles have enough days to accumulate the visiting data, the articles we selected are published during the period from January 21, 2016 to February 18, 2016. It is necessary to note that three articles (1627, 1696 and 1712) are excluded because of the abnormal data. Finally, there are 110 articles selected as our research objects. In summary, visiting data of 110 articles in 90 days are collected and updated daily in this study. The daily harvested visiting data of each articles are parsed into SQL Server database to make analysis.

## Results

*Visiting trend*

Figure 2 illustrates the temporal trend of the daily updated accumulated total visits since the publishing of each paper. Each curve represents for one paper. The research period of all the sample articles is from January 22, 2016 to April 20, 2016, so all the curves end on April 20. For the papers published on the same day, the curves start at the location corresponding to the same day of x axis but with different value of y axis. However, for the papers published on different days, the curves start on different days, so the starting days of the papers are different, when all the starting values of the curves are highlighted with red dots.

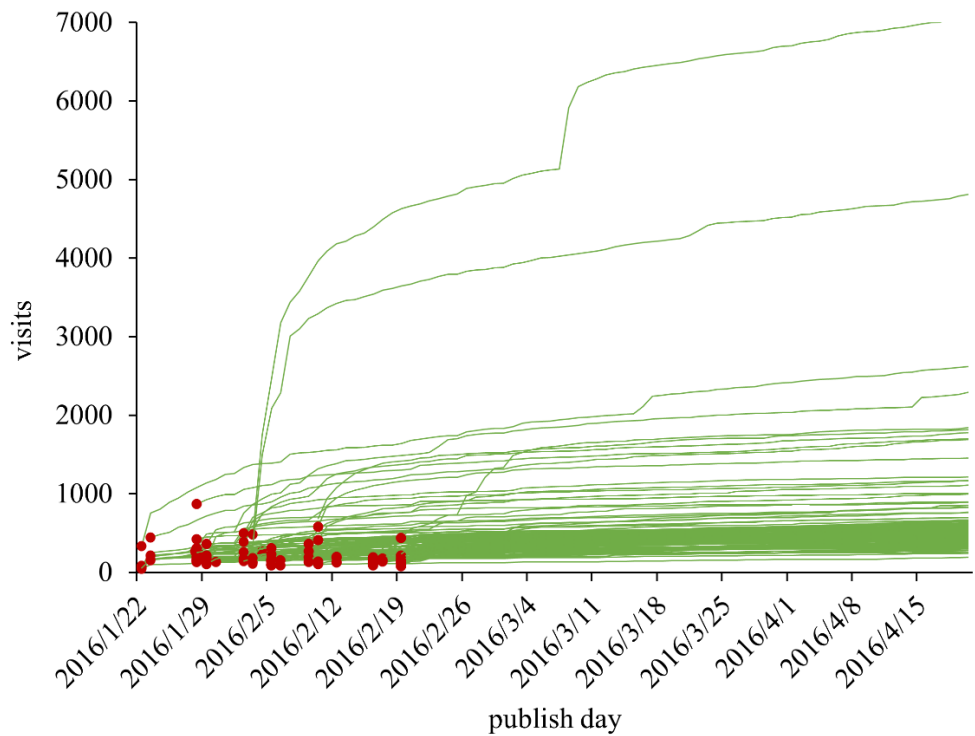

**Fig. 2** Temporal trend of the daily total visits to 110 sample articles

As of the end of 90-days research period, the total visits of articles range from 349 to 7004, while most articles have total visits less than 1000, and most of the visits are received in the initial few days. For most of the papers, the growth of accumulated visits shows very slowly upward trend after the initial rapid growth period. However, there are some big jumps occurring in a few curves, e.g., paper 1605, 1657, etc.

Figure 3 shows the temporal trend of daily updated accumulated visits from social media since the publishing of each paper. The steep starting sections of the curves indicate that most visits come quickly in the few days after an article was published, however, the followed rest flat sections of the curves indicate rapid decay for the visits in the following days. Most papers attract visits from social media only on the first day and the next day, as Fig. 3 shows, most of the height of red dots are very close to the maximum value of the curves. Only a few papers continue to receive visits directed by social media. For several papers, there are some sudden jumps for the daily visits directed by social media.

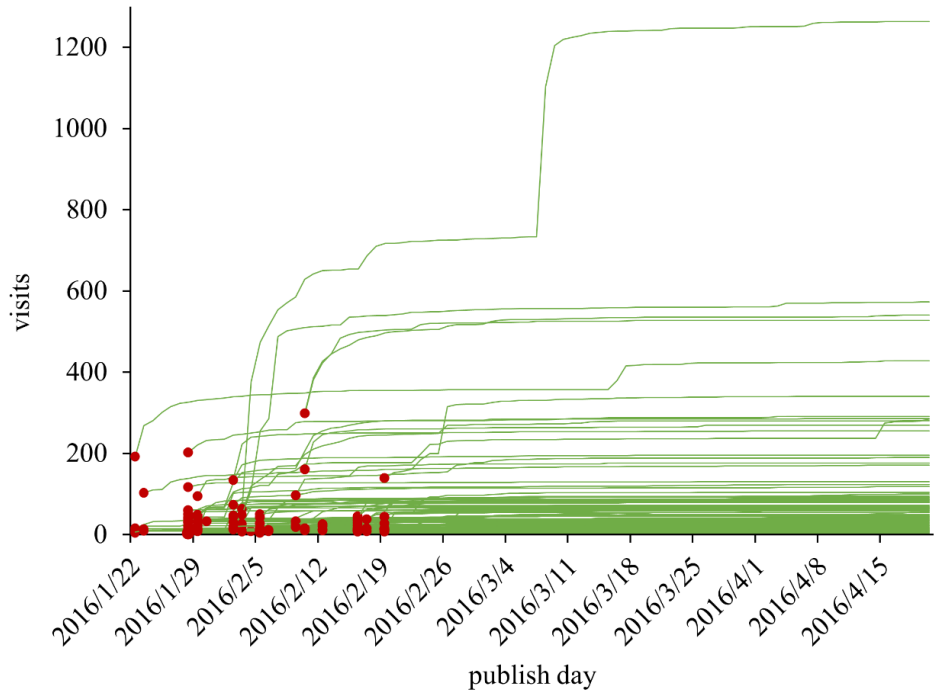

**Fig. 3** Temporal trend of the daily visits directed by social media to 110 sample articles

*Statistics of visits on two days*

We choose the data harvested on two days, which are February 19 and April 20, 2016. February 19 is the first day that all the 110 sample articles have visiting data, including many newly published papers (less than one week) and many papers published for a relative long time (more than 15 days); when April 20 is the last day in the research period and all the papers have been published for at least 60 days.

As Table 1 shows, for the data harvested on February 19, 2016, there are 11 papers published on February 18 with one published day, and 9 papers published on January 25 with 25 published days. In all, there are 26 papers have been published less than one week, 29 papers have been published more than one week but less than half a month, and 44 papers have been published more than half a month but less than one month. However, for the data harvested on April 20, 2016, all the papers have been published for more than 60 days, and we think the difference between 62 days and 86 days is not much and could be accepted in this study.

**Table 1** Date interval from publish date to harvest date of articles

| Publish date | Published days | Number of papers | Harvest date |
| --- | --- | --- | --- |
| 2016/2/18 | 1 | 11 | 2016/2/19 |
| 2016/2/17 | 2 | 4 | 2016/2/19 |
| 2016/2/16 | 3 | 1 | 2016/2/19 |
| 2016/2/15 | 4 | 10 | 2016/2/19 |
| 2016/2/11 | 8 | 5 | 2016/2/19 |
| 2016/2/9 | 10 | 5 | 2016/2/19 |
| 2016/2/8 | 11 | 4 | 2016/2/19 |
| 2016/2/4 | 15 | 15 | 2016/2/19 |
| 2016/2/2 | 17 | 6 | 2016/2/19 |
| 2016/2/1 | 18 | 9 | 2016/2/19 |
| 2016/1/28 | 22 | 10 | 2016/2/19 |
| 2016/1/26 | 24 | 10 | 2016/2/19 |
| 2016/1/25 | 25 | 9 | 2016/2/19 |
| 2016/2/18 | 62 | 11 | 2016/4/20 |
| 2016/2/17 | 63 | 4 | 2016/4/20 |
| 2016/2/16 | 64 | 1 | 2016/4/20 |
| 2016/2/15 | 65 | 10 | 2016/4/20 |
| 2016/2/11 | 69 | 5 | 2016/4/20 |
| 2016/2/9 | 71 | 5 | 2016/4/20 |
| 2016/2/8 | 72 | 4 | 2016/4/20 |
| 2016/2/4 | 76 | 15 | 2016/4/20 |
| 2016/2/2 | 78 | 6 | 2016/4/20 |
| 2016/2/1 | 79 | 9 | 2016/4/20 |
| 2016/1/28 | 83 | 10 | 2016/4/20 |
| 2016/1/26 | 85 | 10 | 2016/4/20 |
| 2016/1/25 | 86 | 9 | 2016/4/20 |

The results are shown in Fig. 4. On February 19, the total visits to 110 articles reach 52630, and 8109 are directed by social referrals, when the proportion of social referrals is about 15.41%. 8109 visits from social media include 4054 from Twitter (50.00%), 3659 from Facebook (45.12%) and 396 from other social media platforms (4.88%), including Google+, Reddit and LinkedIn. When two months later, on April 20, 2016, the total visits to the 110 articles increase to 79946 (increased by 51.90%), and visits from social media increase to 9672 (increased by 19.27%). The visits from social media include 5206 from Twitter (53.83%), 4124 from Facebook (42.64%) and 342 from other social media (3.53%).

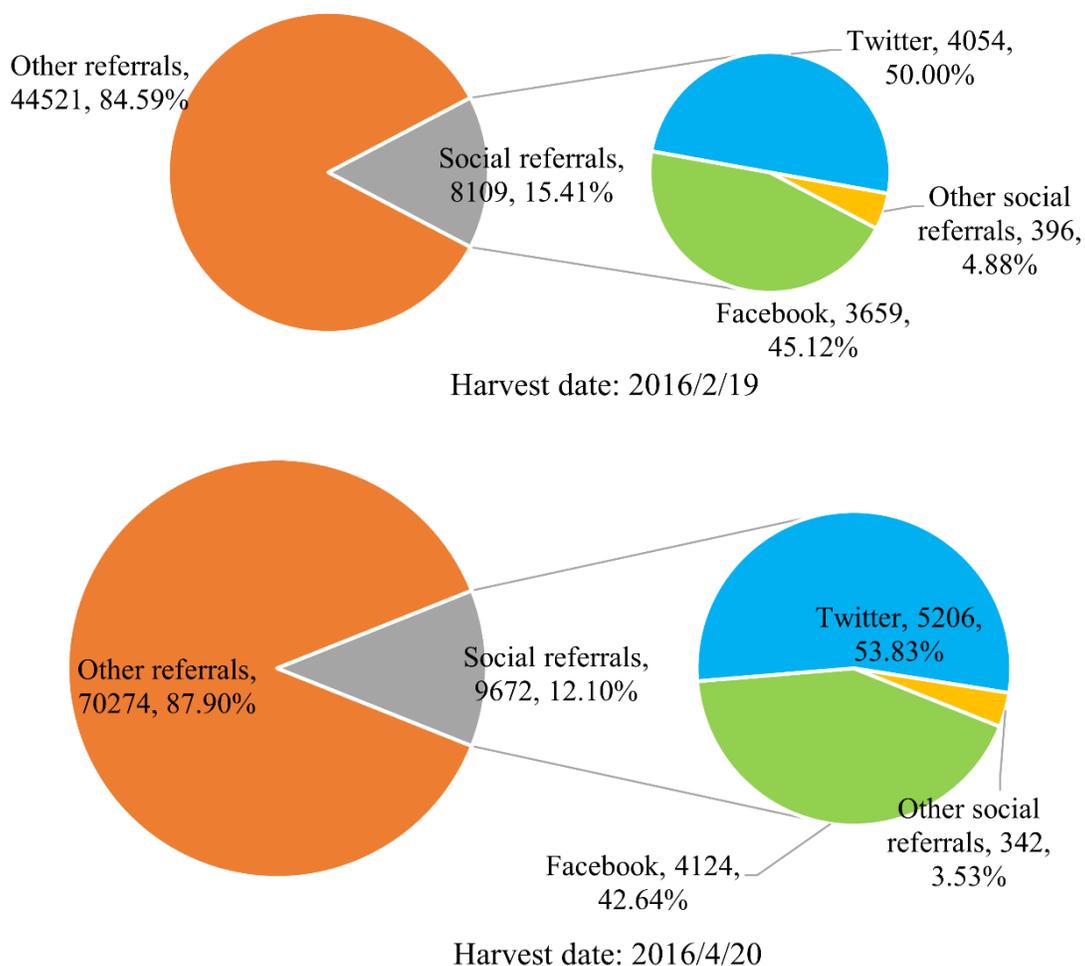

**Fig. 4** The proportion of article visits from social referrals on Feb. 19 and Apr. 20

*Statistics of visits during two periods after publication*

To explore the dynamic changes of the visiting patterns from social media, we process the data as follows, firstly, for each paper on each day from January 22 to April 20, we calculate the publish-harvest interval days between the publish date of articles and harvest date of visiting data. For example, for the paper 1583 (https://doi.org/10.7717/peerj.1583) which is published on January 21, 2016, the interval day on January 22 is one, when the interval days on April 20 are 90. However, for the paper 1686 (https://peerj.com/articles/1686) which is published on February 15, 2016, the interval day on February 16 is one, and the interval days on April 20 are 65. Secondly, all the data are grouped according to the interval days. For example, data with the seven interval days include the data harvested on January 28 of 11 papers published on January 21, the data harvested on February 1 of two papers published on January 25, and so on; when data with 60 interval days include the data harvested on March 21 of 11 papers published on January 21, the data harvested on March 28 of nine papers published on January 28, and so on.

Figure 5 compares the visiting data to papers with seven interval days (left column, seven days after publication) and 60 interval days (right column, 60 days after publication). For the dataset with seven interval days, the total visits are 46,239, among which 7320 visits are from social media and account for 15.83% of the total. When for the dataset with 60 interval days, the total visits reach 74,568 (increased by 61.27%), and visits from social media increase to 9521 (increased by 30.07%, only half of the increase rate of the total visits). And, the 7320 visits from social media in seven days after publication account for 76.88% of the total 9521 visits from social media in 60 days after publication, which means that most of the visits from social media are generated within only one week after publication.

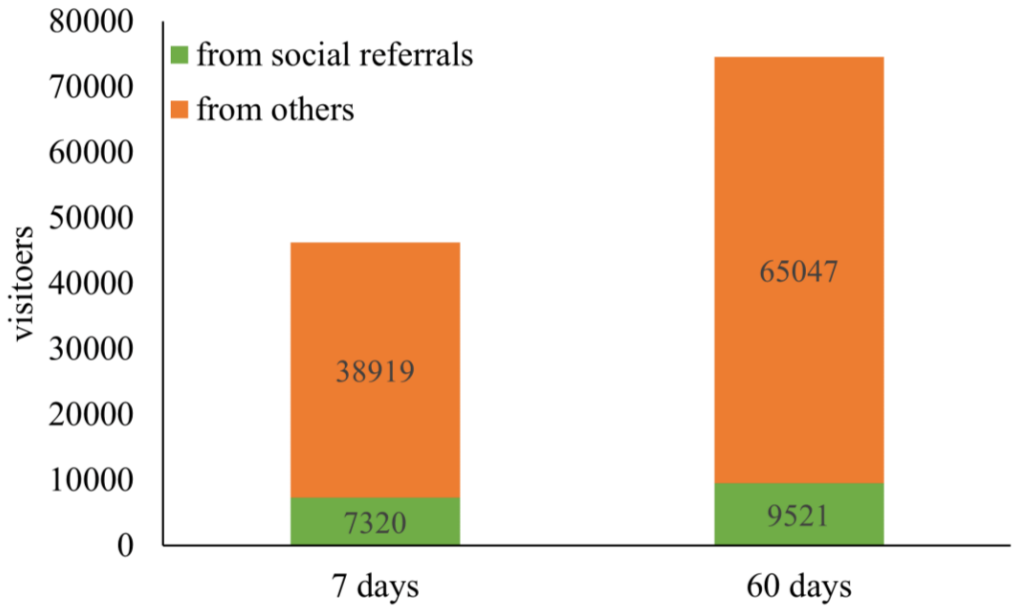

**Fig. 5** Comparison between the visits to papers published 7 and 60 days

Obviously, the increase of visits from social media is not as fast as the increase of the total visits. In another word, social attention to scholarly papers comes in the short term after publication, however, the attention doesn't last long.

In Table 2, we list the percentage of accumulated visits in 7 days of the total accumulated visits in 60 days. For the visits from social media, the median value is 94.47% and the mean value is 87.07%, both of which are greater than the corresponding value of the total visits.

Table 2 Percentage of accumulated visits in 7 days of visits in 60 days

|  | Visits from social media | Total visits |
| --- | --- | --- |
| Max | 100% | 84.44% |
| Min | 10.87% | 40.35% |
| Median | 94.47% | 63.59% |
| Mean | 87.07% | 62.97% |

Figure 6 also shows the box plot of the percentage of social media and total visits. Although the minimum value of the percentage of visits from social media is only 10.87%, the median value of the percentage of social media is much greater than the percentage of total visits. The left box plot is much higher than the right box plot, which suggests that the visits from social media are much faster to accumulate than visits from other referrals.

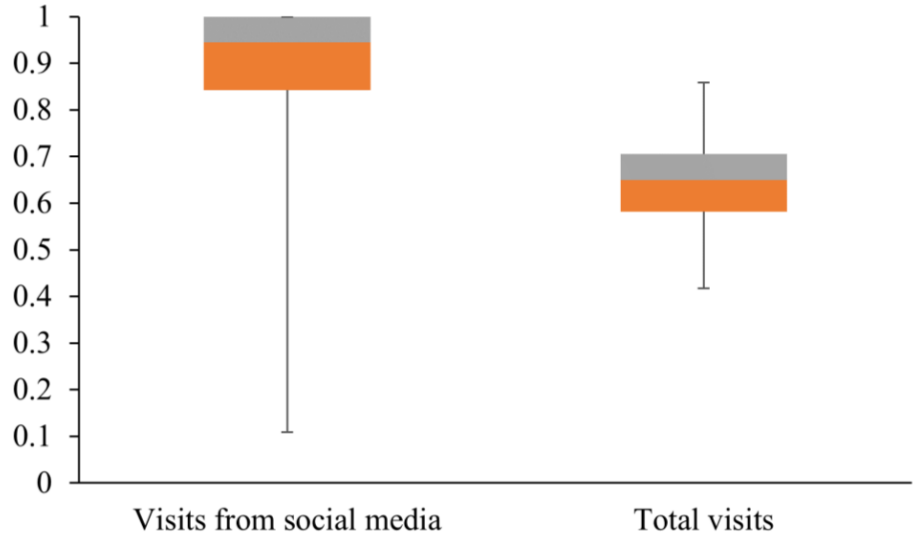

**Fig. 6** Box plot of percentage of accumulated visits in 7 days of visits in 60 days

*Visiting trend*

To track the temporal trend of the changes of visits from social media, the daily harvested visiting data of each article are grouped according to the publish-harvest interval days, as mentioned above. The results are shown in Fig. 7. The papers which have been published for one day have the greatest percentage, visits from social media account for 20.28% of the total visits. For the papers with 90 interval days, the percentage decreases to 9.06%. The curve in Fig. 7 shows obvious overall downward trend.

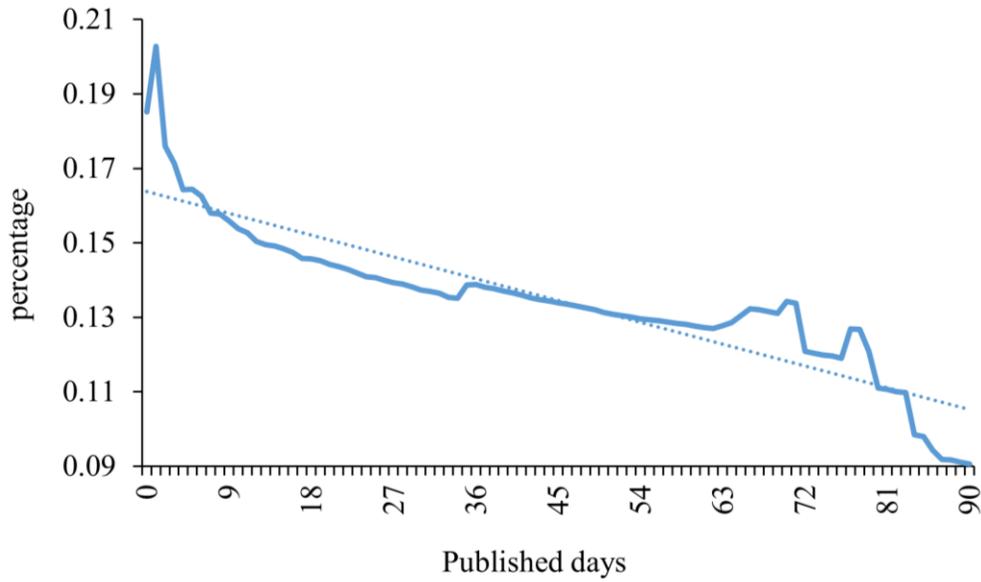

**Fig. 7** Temporal trend of the proportion of visits from social media in the total visits

A burst occurs at the middle part of the curve, we check the data and find out the reason, the burst is resulted by a jump for the visits from social media of the paper 1605 (https://doi.org/10.7717/peerj.1605) on March 8, 2016, when the paper has been published for 35 days, its visits from social media increase from 733 to 1103 and increased by 50.48%. On the next day, the visits from social media continue to increase to 1204 and keep stable from then on.

The fluctuation of the tail of the curve is caused by the decrease of the sample articles. In our dataset, all the sample articles are published before February 18, and have the publish-harvest interval days greater than 60, however, there are only 84 papers have the interval days greater than 65, which could explain why the fluctuation begins from the 66 interval days of the curve.

*Visits from social media: case study of paper 1605*

As discussed above, there is a spike on the daily visits from social media for paper 1605, here we use the data of paper 1605 to make more detailed analysis of the temporal trends.

The visiting data directed by Twitter are extracted from the dataset, the tweets about the article and twitter users' followers are checked and retrieved directly from

twitter.com. So, three kinds of data are used to examine the relationship between twitter and the resulted visits to scholarly articles, which are the number of tweets about an article, the followers of the related twitter accounts, and the number of article visitors directed by the tweets. The paper 1605 was published on February 2, 2016, on the next day, it has been tweeted by 63 Twitter accounts with 118,151 followers, and the number of article visitors directed by Twitter reached 39. On February 4, article visitors rose dramatically to 204. On February 8, the number of Twitter accounts reached 72 with 138,164 followers, while the number of article visitors reached 309. From then to March 6, the number of Twitter accounts rose from 72 to 77, and the resulted article visitors rose from 309 to 381 correspondingly. On March 7, an influential Twitter account with 1.97 million followers tweeted the paper, which has been retweeted 11 times on the same day. This is the reason why the number of article visitors from Twitter rose dramatically from 381 to 751 simultaneously, which explains the sudden jumps for both the two curves in Fig. 8 and 9.

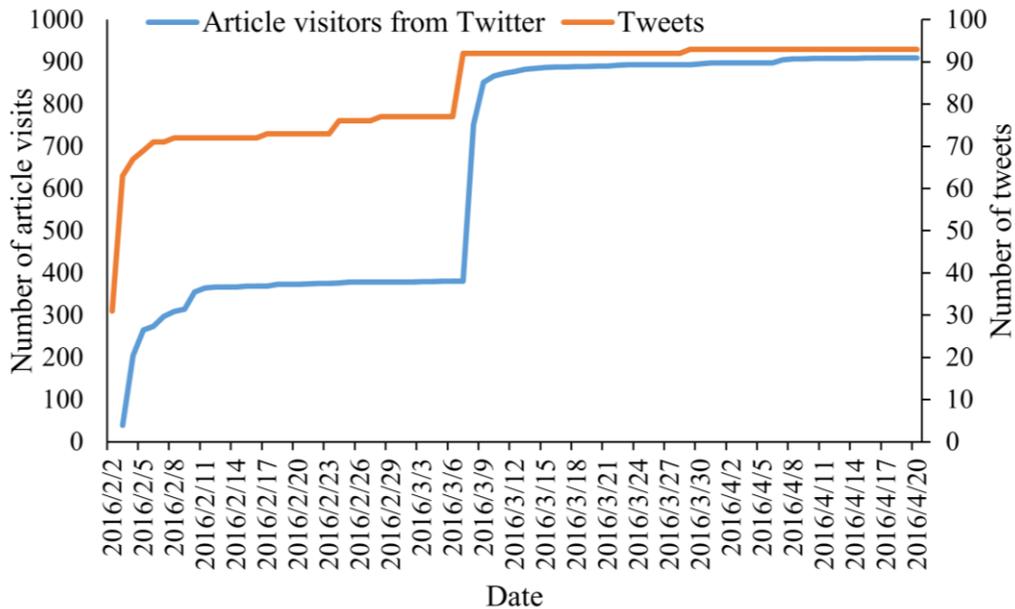

**Fig. 8** Synchronism of temporal trend of tweets and their resulted visits for paper 1605

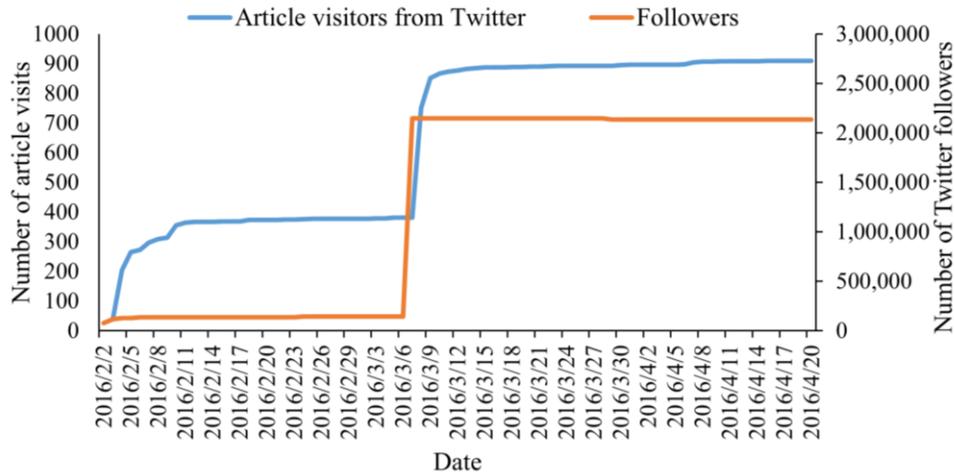

**Fig. 9** Synchronism of temporal trend of the article visits from Twitter and followers of related Twitter accounts for paper 1605

The synchronism of the growth of tweets and tweet-resulted article visitors testifies partially that social mention directs people to read scholarly articles, although we don't know who are directed by which tweet. Article visitors from social referrals may be researchers, students, or even general public.

**Conclusion and discussion**

Social media plays an important role in directing people to scholarly articles, which is confirmed for the first time with real data in this study. According to our empirical study with 110 sample articles, article visits directed by social referrals account for about 12.10% (final period) to 15.41% (initial period) of total visits. At the initial period after the publication of scholarly article, social attention comes very quickly. From the temporal perspective, the distribution of visits from social media is very uneven. In most cases, visits from social media are much faster to accumulate than visits from other referrals, most visits directed by social referrals are concentrated in the initial few days after publication. According to our statistics, 76.88% of the visits from social media are generated in the initial week after publication. Easy come, easy go, social buzz about scholarly articles doesn't last long, which leads to the resulted article visits a rapid decay.

Twitter and Facebook are the two most important social referrals that directing people to scholarly articles, the two are about the same and account for over 95% of the total social referral directed visits. And the share of other social media is negligible, including LinkedIn, Google+, Reddit, etc.

The visiting dynamics analysis also shows that there is an obvious overall downward temporal trend for the percentage of visits from social media in total visits. For papers one day after publication, social referral directed visits account for 20.28% of all visits. After 90 days, the percentage decreases to only 9.06%.

For some papers, there are spikes for the daily social referral directed visits. However, the few spikes have great influences on the overall situation. Taking paper 1605 as the case, we find that the spike of visits is caused by the spike of tweets, there is synchronism between tweets and tweets resulted visits.

There are some limitations for this study. Because only *PeerJ* provides the referral data of each article to general public, our analysis is based on 110 sample articles published by *PeerJ* since January 22, 2016, the results and conclusions are expecting more evidences from other journals with the availability of referral data in the future.

## Acknowledgements

The work was supported by the project of ''National Natural Science Foundation of China'' (61301227), the project of "the Fundamental Research Funds for the Central Universities"(DUT15YQ111), "Liaoning Province Higher Education Innovation Team Fund" (WT2015002), and "Humanity and Social Science Foundation of Ministry of Education of China'(11YJA630128) .